\documentclass[10pt,conference]{IEEEtran}

\usepackage{multicol}
\usepackage{graphics}

\newtheorem{theorem}{Theorem}[section]

\newcommand{\bi}{\begin{itemize}}
\newcommand{\ei}{\end{itemize}}
\newcommand{\ben}{\begin{enumerate}}
\newcommand{\een}{\end{enumerate}}
\newcommand{\beq}{\begin{equation}}
\newcommand{\eeq}{\end{equation}}
\newcommand{\beqa}{\begin{eqnarray}}
\newcommand{\eeqa}{\end{eqnarray}}

\begin{document}

\title{Tree-Based Construction of LDPC Codes}

\author{\authorblockN{Deepak Sridhara,  Christine Kelley, and  Joachim Rosenthal$^{1}$\footnotemark}\authorblockA{Institut f$\ddot{\mbox{u}}$r Mathematik,\\
Universit$\ddot{\mbox{a}}$t Z$\ddot{\mbox{u}}$rich,\\
CH-8057 Z$\ddot{\mbox{u}}$rich, Switzerland.\\
email: \{cak, rosen, sridhara\}@math.unizh.ch}
 }

\maketitle
\footnotetext[1]{This work was supported by NSF Grant No. 
CCR-ITR-02-05310.}

\begin{abstract}
We present a construction of LDPC codes that have minimum pseudocodeword weight equal to the minimum distance, and perform well with iterative decoding. The construction involves enumerating a $d$-regular tree for a fixed number of layers and employing a connection algorithm based on mutually orthogonal Latin squares to close the tree. Methods are presented for degrees $d=p^s$ and $d = p^s+1$, for $p$ a prime, -- one of which includes the well-known finite-geometry-based LDPC codes.
\end{abstract}
\section{Introduction}

Low Density Parity Check (LDPC) codes are widely acknowledged to be good 
codes due to their near Shannon-limit performance when decoded 
iteratively. However, many structure-based constructions of LDPC codes 
fail to achieve this level of performance, and are often outperformed by 
random constructions.  (Exceptions include the finite-geometry-based LDPC 
codes (FG-LDPC) of \cite{kou}, which were later generalized in \cite{xu}.) 
Moreover, there are discrepancies between iterative and maximum likelihood 
(ML) decoding performance of short to moderate blocklength LDPC codes. 
This behavior has recently been attributed to the presence of so-called 
{\em pseudocodewords} of the LDPC constraint graphs, which are valid 
solutions of the iterative decoder which may or may not be optimal 
\cite{pascal}. Analogous to the role of minimum Hamming distance, 
$d_{\min}$, in ML-decoding, the minimal pseudocodeword weight, $w_{\min}$, 
has been shown to be a leading predictor of performance in iterative 
decoding. Furthermore, the error floor performance of iterative decoding 
is dominated by minimal weight pseudocodewords. Although there exist 
pseudocodewords with weight larger than $d_{\min}$ that have adverse 
affects on decoding, pseudocodewords with weight $w_{\min} < d_{\min}$ are 
especially problematic \cite{jpb}.

The Type I-A construction and certain cases of the Type II construction 
presented in this paper are designed so that the resulting codes have 
minimal pseudocodeword weight equal to the minimum distance of the code, 
and consequently, these problematic low-weight pseudocodewords are 
avoided.  The resulting codes have minimum distance which meets the lower 
tree bound originally presented in \cite{tanner}, and since $w_{\min}$ 
shares the same lower bound \cite{jpb,isit2004}, and is upper bounded by 
$d_{\min}$, the proposed constructions have $w_{\min} = d_{\min}$. 
It is worth noting that this property is also a characteristic of some of 
the FG -LDPC codes \cite{xu}, and indeed, the projective-geometry-based 
codes of \cite{kou} arise as special cases of our Type II construction. 
Furthermore, the Type I-B construction presented herein is a modification 
of the Type I-A construction, and it yields a family of codes with a wide 
range of rates and blocklengths that are comparable to those obtained from 
finite geometries.

We now present the tree bound on $w_{\min}$  derived in \cite{isit2004}.

\begin{theorem}
{ \em Let $G$ be a bipartite LDPC constraint graph with smallest left (variable node) degree $d$ and girth $g$. Then the minimal pseudocodeword weight $w_{\min}$ (for the AWGN/BSC channels) is lower bounded by
\vspace{-0.1in}
 \[\scriptsize w_{\min} \ge \left\{\begin{array}{cc}
1+d + d(d-1) + d(d-1)^2+\ldots+ d(d-1)^{\frac{g-6}{4}}, & \frac{g}{2}\mbox{ odd }\\
1+ d+d(d-1)+\ldots+d(d-1)^{\frac{g-8}{4}} + (d-1)^{\frac{g-4}{4}},& \frac{g}{2}\mbox{ even }
\end{array}\right . \]
}
\label{thm1}\vspace{-0.15in}
\end{theorem}

This bound is also the tree bound  on the minimum distance established by Tanner in \cite{tanner}. And since the set of pseudocodewords includes all codewords, we have $w_{\min}\le d_{\min}$. In the following sections we present two construction techniques of LDPC codes wherein for certain cases, $w_{\min} = d_{\min}$.

\section{preliminaries}

 The connection algorithms for the tree constructions Type I-B and Type II 
are based on mutually orthogonal Latin squares. A well-known construction 
of a family of mutually orthogonal Latin squares of order $p^s$, a power 
of a prime, may be found in \cite{roberts}. Let 
$M^{(1)},M^{(2)},\dots,M^{(p^s-1)}$ denote $p^s-1$ mutually orthogonal 
Latin squares (MOLS) of order $p^s$. Let the rows (and columns) of each 
square be indexed by the integers $0,1,2,\dots,p^s-1$. Without loss of 
generality, assume that the first column of each of the Latin squares is 
$[0,1,2,\dots,p^s-1]^T$.  In addition, define a new square of size 
$p^s\times p^s$, denoted $M^{(0)}$, where each column of $M^{(0)}$ 
is $[0,1,2,\dots,p^s-1]^T$.

 \section{Tree-based Construction: Type I}
In the Type I construction, first a $d$-regular tree of alternating variable and constraint node layers is enumerated from a root variable node (layer $ L_0$) for $\ell$ layers. If $\ell$ is odd (respectively, even), the final layer $L_{\ell - 1}$ is composed of variable nodes (respectively, constraint nodes). Call this tree $T$. The tree $T$ is then reflected across an imaginary horizontal axis to yield another tree, $T'$,  and  the variable and constraint nodes are reversed. That is, if layer $L_i$ in $T$ is composed of variable nodes, then the reflection of $L_i$, call it $L_i'$, is composed of constraint nodes in the reflected tree, $T'$. The union of these two trees, along with edges connecting the nodes in layers $L_{\ell-1}$ and $L_{\ell - 1}'$ according to a connection algorithm that is described next, comprise the graph representing a Type I LDPC code. We now present two connection schemes that can be used in this Type I model, and discuss the resulting codes.
\subsection{Type I-A}
For $d=3$, the Type I-A construction yields a $d$-regular LDPC constraint graph  having $1+d+d(d-1)+\ldots+d(d-1)^{\frac{g-4}{2}}$ variable and constraint nodes, and girth $g$.  The tree $T$ has $\frac{g}{2}$ layers. To connect the nodes in $L_{\frac{g}{2} - 1}$ to $L_{\frac{g}{2} - 1}'$, first label the variable (resp., constraint) nodes in $L_{\frac{g}{2}-1}$ (resp., $L_{\frac{g}{2}-1}'$) when $\frac{g}{2}$ is odd, as $v_0,v_1,\dots,v_{2^{\frac{g}{2}-2}-1}$, $v_{2^{\frac{g}{2}-2}},\dots,v_{2\cdot 2^{\frac{g}{2}-2}-1},v_{2\cdot 2^{\frac{g}{2}-2}},\dots,v_{3\cdot 2^{\frac{g}{2}-2}-1}$ (resp., $c_0,c_1,\dots,c_{3\cdot 2^{\frac{g}{2}-2}-1}$). The nodes $v_0,v_1,\dots,v_{2^{\frac{g}{2}-2}-1}$ form the $0^{th}$ class, the nodes $v_{2^{\frac{g}{2}-2}},\dots,v_{2\cdot 2^{\frac{g}{2}-2}-1}$ form the $1^{st}$ class, and the nodes $v_{2\cdot 2^{\frac{g}{2}-2}},\dots,v_{3\cdot 2^{\frac{g}{2}-2}-1}$ form the $2^{nd}$ class; classify the constraint nodes in a similar manner. In addition, define three permutations $\pi(\cdot), \tau(\cdot),\tau'(\cdot)$ of the set $\{0,1,\dots,2^{\frac{g}{2}-2}-1\}$ as follows.
The nodes in $L_{\frac{g}{2} - 1}$ and $L_{\frac{g}{2} - 1}'$ are connected as follows:
\begin{enumerate}
\item For $i=0,1$, and $j=0,1,\dots,2^{\frac{g}{2}-2}-1$, the variable node $v_{j+i\cdot 2^{\frac{g}{2}-2}}$ is connected to nodes $c_{\pi(j)+i\cdot 2^{\frac{g}{2}-2}}$ and $c_{\tau(j)+(i+1)\cdot 2^{\frac{g}{2}-2}}$.
\item For $i=2$, and $j=0,1,\dots,2^{\frac{g}{2}-2}-1$, the variable node $v_{j+i\cdot 2^{\frac{g}{2}-2}}$ is connected to nodes $c_{\pi(j)+2\cdot 2^{\frac{g}{2}-2}}$ and $c_{\tau'(j)}$.
\end{enumerate}
The permutations for the cases $g=6,8,10,12$ are given below. The above construction can be extended for higher $g$ in a natural way and we are working on an explicit closed form expression for the permutations $\pi, \tau, \tau'$ for higher $g$.

\begin{figure}[t]
 \centering{\resizebox{3.25in}{1.475in}{\includegraphics{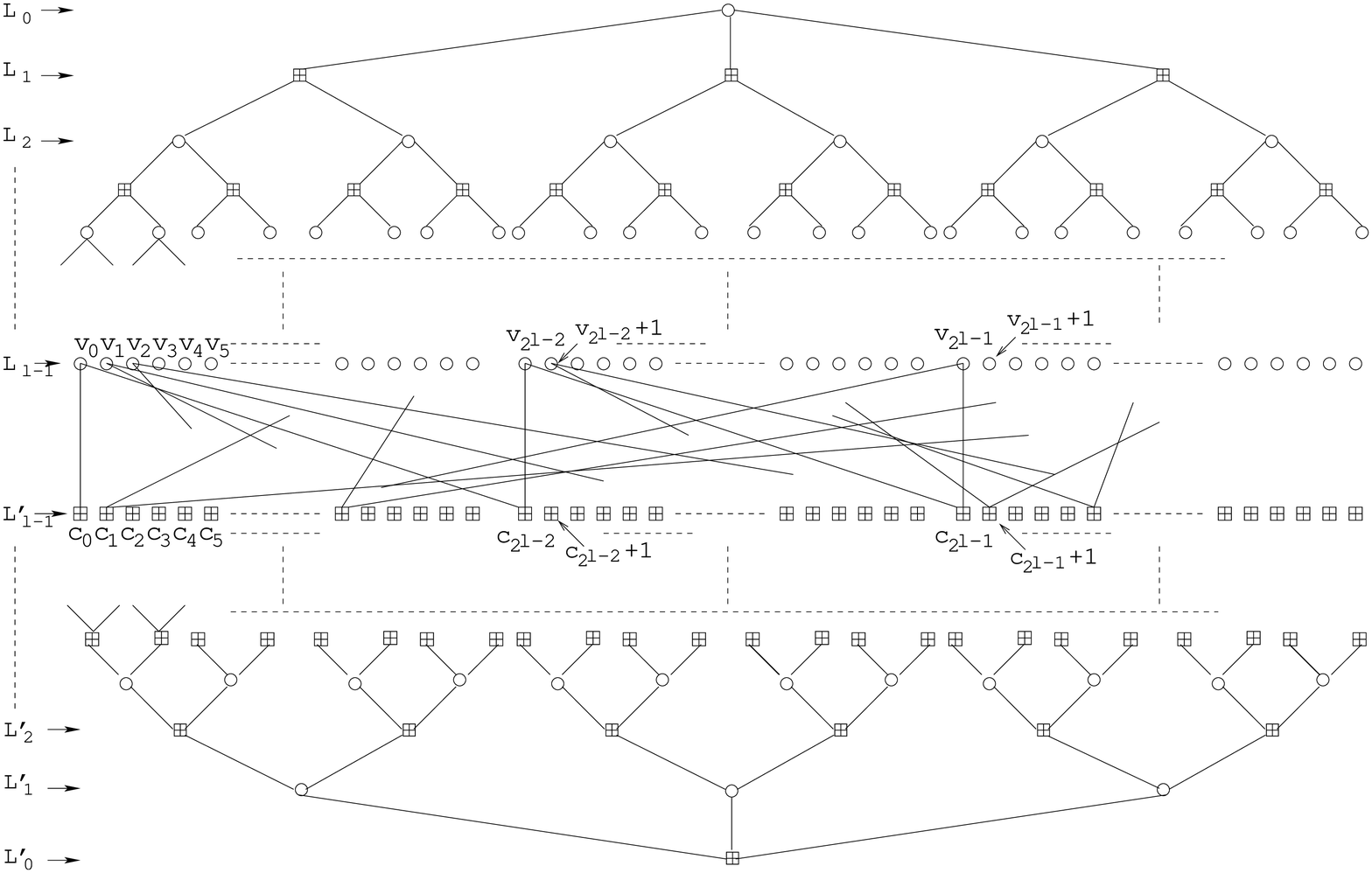}}}
\caption{Tree construction of Type I-A LDPC code. }
\label{type1A_const}
\end{figure}

\begin{figure}[t]
 \centering{\resizebox{3.25in}{1.475in}{\includegraphics{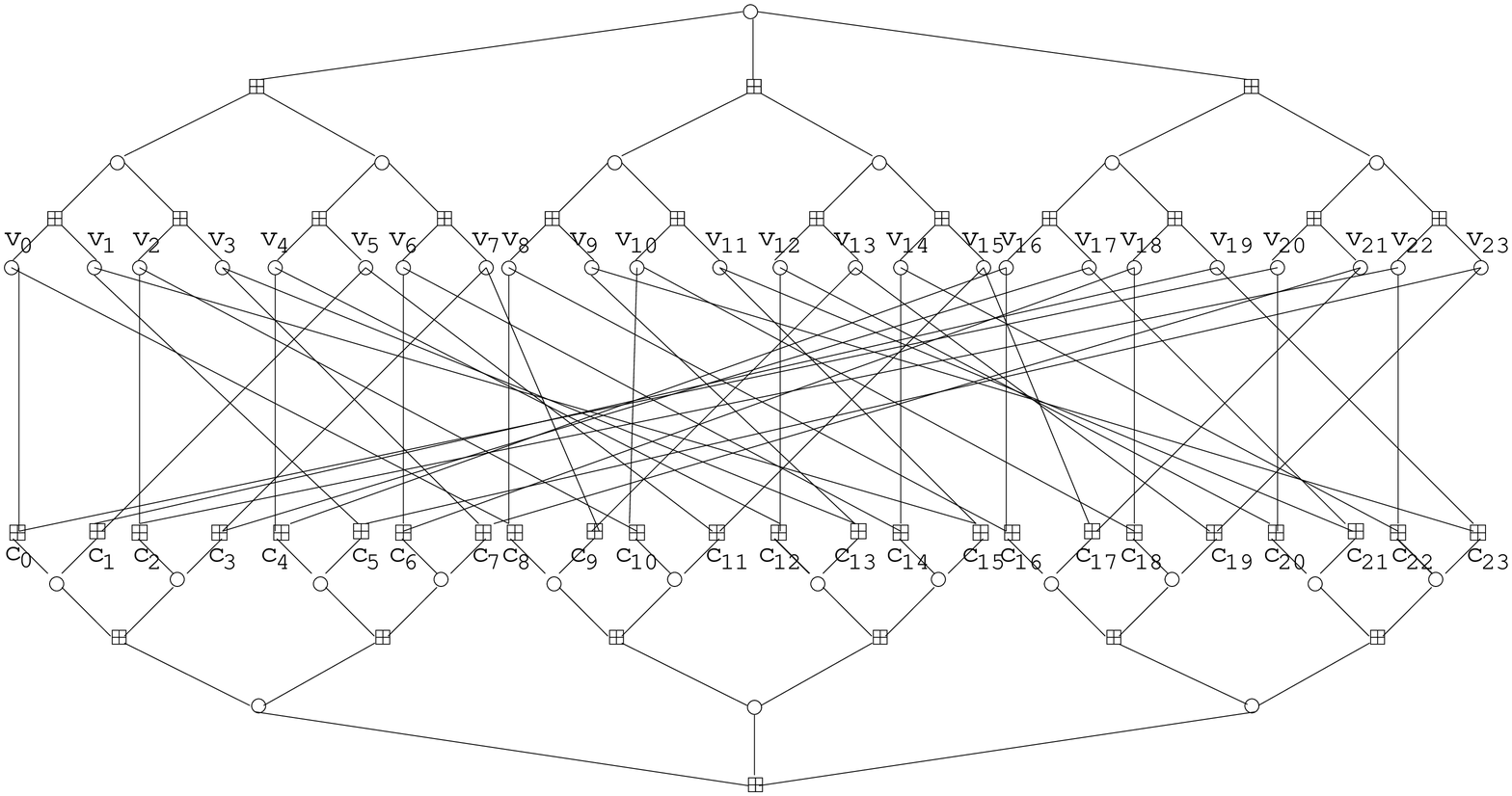}}}
\caption{Type I-A LDPC constraint graph having degree $d=3$ and girth $g=10$.}
\label{type1Ad3g10_graph}
\end{figure}

\vspace{-0.1in}
{\scriptsize
\[g=6, \pi= \tau=\tau'=(0)(1), \mbox{the identity permutation.} \]
\vspace{-0.1in}\[g=8, \pi=(0)(2)(1,3), \tau=(0)(2)(1,3), \tau'=(0,2)(1)(3).  \]
\vspace{-0.1in}\[\hspace{-0in}g=10,\ \pi=(0)(2)(4)(6)(1,5)(3,7), \tau=(0)(2)(4)(6)(1,7)(3,5),\]
\vspace{-0.1in}\[\tau'=(0,4)(2,6)(1,3)(5,7).\]
\vspace{-0.1in}\[ g=12, \pi=(0)(4)(8)(12)(2,6)(10,14)(1,9)(3,15)(5,13)(7,11), \]
\vspace{-0.1in}\[\tau=(0)(4,12)(8)(2,6,10,14)(1,15,13,11)(3,9,7,5), \]
\vspace{-0.1in}\[\tau'=(0,8)(4,12)(2,14)(6,10)(1,3,5,7)(9,11,13,15). \]
}
When $\frac{g}{2}$ is odd, the minimum distance of the resulting code meets the tree bound, and hence, $d_{\min} = w_{\min}$. When $\frac{g}{2}$ is even, $d_{\min}$ is strictly larger than the tree bound; we believe however, that  $w_{\min}$ is equal to $d_{\min}$ in this case as well. Figure~\ref{type1A_const} illustrates the general construction procedure and Figure~\ref{type1Ad3g10_graph}  shows a girth 10 Type I-A LDPC constraint graph.


\subsection{Type I-B}
\begin{figure}[t]
\centering{\resizebox{3.25in}{1.6in}{\includegraphics{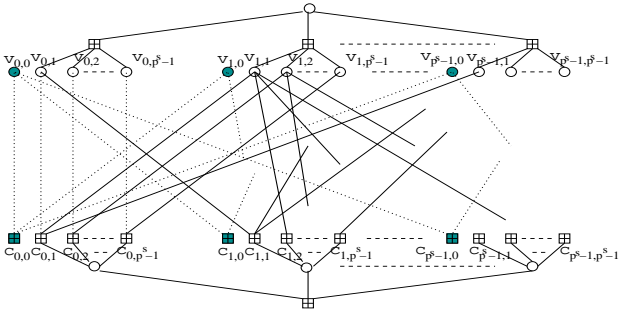}}}
\caption{Tree construction of Type I-B LDPC code. (Shaded nodes are imaginary nodes and dotted lines are imaginary lines.)}
\label{type1B_const}
\end{figure}

\begin{figure}[t]
\centering{\resizebox{3.25in}{1.6in}{\includegraphics{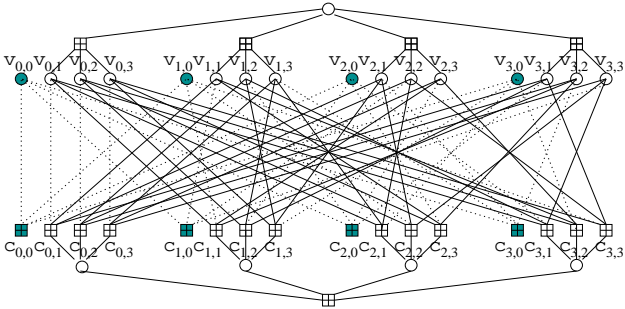}}}
\caption{Type I-B LDPC constraint graph having degree $d=4$ and girth $g=6$.}
\label{type1Bd4g6_graph}
\end{figure}

For $d = p^s, p$ a prime, the Type I-B construction yields a $d$-regular 
LDPC constraint graph having $1+d+d(d-1)$ variable and constraint nodes, 
and girth $6$. The tree $T$ has 3 layers $L_0,L_1,$ and $L_2$. $L_2$ 
(resp., $L_2'$) is composed of $p^s$ sets $\{S_i\}_{i=0}^{p^s-1}$ 
of $p^s - 1$ variable (resp., constraint) nodes in each set; the 
set $S_i$ corresponds to the children of branch $i$ of the root node. Let 
$S_i$ (resp., $S_i'$) contain the variable (resp., constraint) nodes 
$v_{i,1},v_{i,2},\ldots,v_{i,p^s-1}$ (resp., 
$c_{i,1},c_{i,2},\ldots,c_{i,p^s-1}$). To use MOLS of order $p^s$ in the 
connection algorithm, an imaginary node, $v_{i,0}$ (resp., $c_{i,0}$) is 
temporarily introduced into each set $S_i$ (resp, $S_i'$). The connection 
algorithm proceeds as follows:

\begin{enumerate}
\item Let $x_t(i,j)$ denote the $(j,t)^{th}$ entry of the square $M^{(i)}$ 
defined in Section II. For $i = 0,\ldots,p^s-1$ and $j = 0,\ldots,p^s-1$, 
connect variable node $v_{i,j}$ to constraint nodes 
$c_{0,x_0(i,j)},c_{1,x_1(i,j)},\ldots,c_{p^s-1,x_{p^s-1}(i,j)}$.
\item Delete all imaginary nodes $\{v_{i,0},c_{i,0}\}_{i=0}^{p^s-1}$ and 
the edges incident on them.
\item For $i = 1,\ldots,p^s-1,$ delete the edge connecting $v_{0,i}$ to 
$c_{0,i}$.
\end{enumerate}
The resulting $d$-regular constraint graph represents the Type I-B LDPC 
code. Figure~\ref{type1B_const} illustrates the construction procedure and 
Figure \ref{type1Bd4g6_graph} provides a specific example of a Type I-B 
LDPC constraint graph with $d=4$; the squares used for constructing this 
graph are

\[\tiny 
\hspace{-0in} \left[\begin{array}{cccc}
0&0&0&0\\
1&1&1&1\\
2&2&2&2\\
3&3&3&3
\end{array}\right], \ \left[\begin{array}{cccc}
0&1&2&3\\
1&0&3&2\\
2&3&0&1\\
3&2&1&0
\end{array}\right], \ \left[\begin{array}{cccc}
0&2&3&1\\
1&3&2&0\\
2&0&1&3\\
3&1&0&2
\end{array}\right], \ \left[\begin{array}{cccc}
0&3&1&2\\
1&2&0&3\\
2&1&3&0\\
3&0&2&1
\end{array}\right] .\]

The Type I-B algorithm yields LDPC codes having a wide range of rates and 
blocklengths that are comparable to, but different from, the 
two-dimensional LDPC codes from 
finite Euclidean geometries \cite{kou,xu}. The Type I-B LDPC
codes are $p^s$-regular with girth six, blocklength $N=p^{2s}+1$,
and distance $d_{\min}\ge p^s+1$. For degrees of the form 
$d=2^s$, the resulting Type I-B codes have very good rates, above
0.5, and perform well with iterative decoding. 


\section{Tree-based Construction: Type II}
In the Type II construction, first a $d$-regular tree $T$ of alternating variable and constraint node layers is enumerated from a root variable node (layer $ L_0$) for $\ell$ layers, as in Type I. The tree $T$ is not reflected; rather, a single layer of $(d-1)^{\ell-1}$ nodes is added to form layer $L_{\ell}$. If $\ell$ is odd (resp., even), this layer is composed of constraint (resp., variable) nodes. The union of $T$ and $L_{\ell}$,  along with edges connecting the nodes in layers $L_{\ell-1}$ and $L_{\ell}$ according to a connection algorithm that is described next, comprise the graph representing a Type II LDPC code. We now present the  connection scheme that is used for this Type II model, and discuss the resulting codes.\\
 The connection algorithm for $\ell = 3$ and $\ell = 4$ proceeds as follows.
\subsection{ $\ell = 3$}
\begin{figure}[t]
\centering{\resizebox{3.25in}{1.6in}{\includegraphics{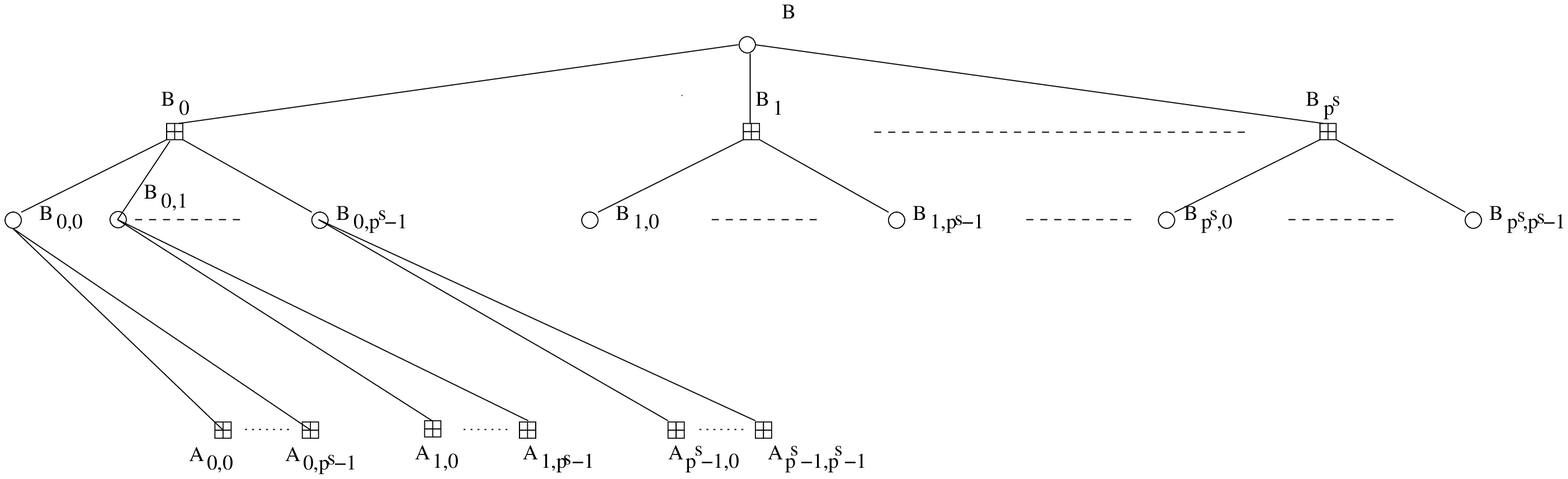}}}
\caption{Tree construction of girth 6 Type II ($\ell=3$) LDPC code.}
\label{type2g6_const}
\end{figure}

\begin{figure}[t]
\centering{\resizebox{3.25in}{1.6in}{\includegraphics{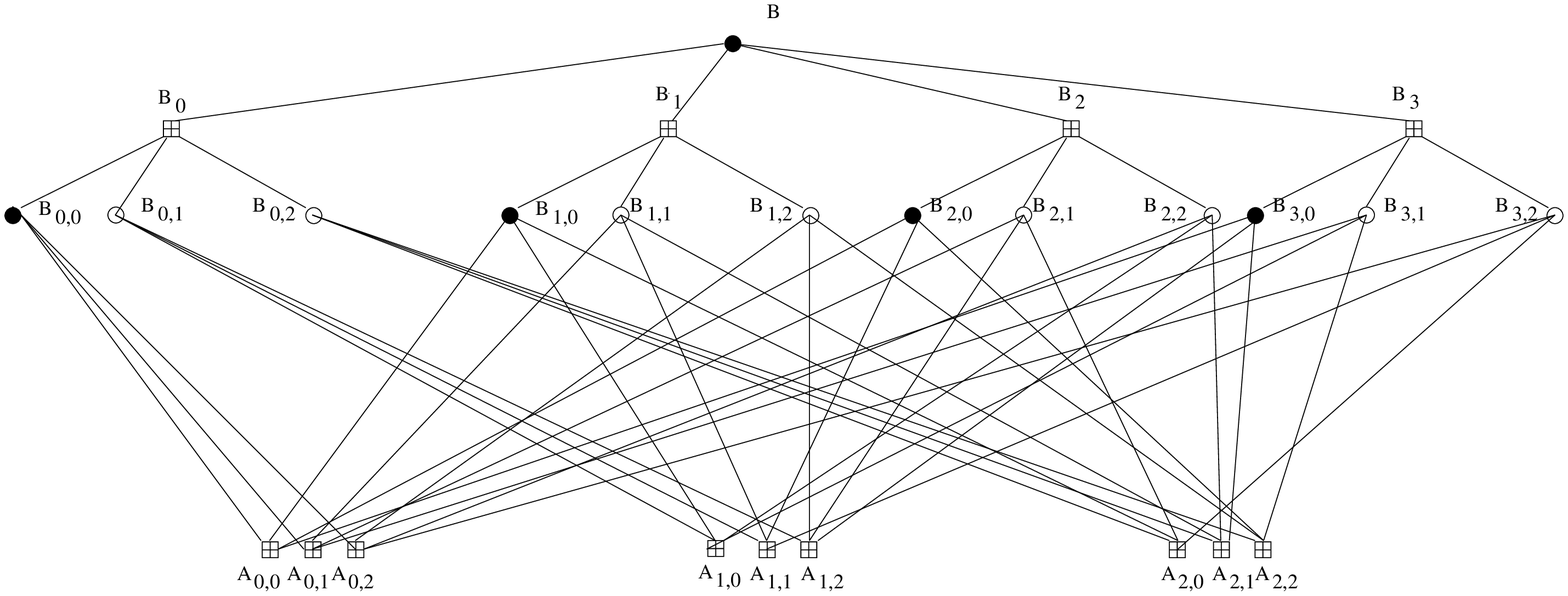}}}
\caption{Type II LDPC constraint graph having degree $d=4$ and girth $g=6$. (Shaded nodes highlight a minimum weight codeword.)}
\label{type2g6d4_graph}
\end{figure}

\indent The $d$ constraint nodes in $L_1$ are labeled $B_0,B_1,\dots,B_{p^s}$ to represent the $d$ branches stemming from the root node, and the $d(d-1)$ variable nodes in the third layer $L_2$ are labeled as $B_{0,0},B_{0,1},\dots,B_{0,p^s-1}$, $B_{1,0}, \dots, B_{1,p^s-1}$, $\dots$, $B_{p^s,0}, \dots,B_{p^s,p^s-1}$.
The $p^{2s}$ constraint nodes in the final layer $L_{\ell}=L_{3}$  are labeled $A_{0,0},A_{0,1},\dots,A_{0,p^s-1}$, $A_{1,0},A_{1,1},\dots,A_{1,p^s-1}$, $\dots$, $A_{p^s-1,0},A_{p^s-1,1},\dots,A_{p^s-1,p^s-1}$.
\begin{enumerate}
\item The constraint nodes in $L_3$ are grouped into $d-1=p^s$ classes of $d-1=p^s$ nodes in each class. Similarly, the variable nodes in $L_2$ are grouped into $d=p^s+1$ classes of $d-1=p^s$ nodes in each class. Those nodes descending from $B_{0}$ form the $0^{th}$ class, those descending from $B_{1}$ form the first class, and so on.
\item Each of the variable nodes descending from $B_{0}$ is connected to all the constraint nodes of one class. That is, $B_{0,0}$ is connected to $A_{0,0}, A_{0,1},\dots,A_{0,p^s-1}$, $B_{0,1}$ is connected to $A_{1,0}, A_{1,1},\dots, A_{1,p^s-1}$, and in general, $B_{0,k}$ is connected to $A_{k,0}, A_{k,1},\dots,A_{k,p^s-1}$ which correspond to the constraint nodes in the $k^{th}$ class.
\item Let $x_t(i,j)$ denote the $(j,t)^{th}$ entry of  $M^{(i-1)}$.  
\item Then connect the variable node $B_{i,j}$ to the constraint nodes\vspace{-0.15in} $$A_{0,x_0(i,j)}, A_{1,x_1(i,j)}, A_{2,x_2(i,j)},\dots, A_{p^s-1,x_{p^s-1}(i,j)}.$$\vspace{-0.15in}
\end{enumerate}\vspace{-0.1in}
Figure~\ref{type2g6_const} illustrates the construction procedure and Figure \ref{type2g6d4_graph} provides an example of a Type II LDPC constraint graph with degree $d=4$ and girth $g=6$; the squares used for constructing this example are
\[\tiny 
\hspace{-0in} M^{(0)}=\left[\begin{array}{ccc}
0&0&0\\
1&1&1\\
2&2&2
\end{array}\right], \ M^{(1)}=  \left[\begin{array}{ccc}
0&1&2\\
1&2&0\\
2&0&1
\end{array}\right], \ M^{(2)}=\left[\begin{array}{ccc}
0&2&1\\
1&0&2\\
2&1&0
\end{array}\right].\]
The ratio of minimum distance to blocklength of the codes is at least $\frac{2+p^s}{1+p^s+p^{2s}}$, and the girth is six. For degrees $d$ of the form $d=2^s+1$, the tree bound on minimum distance and minimum pseudocodeword weight \cite{tanner,isit2004} is met, i.e., $d_{\min}=w_{\min}=2+2^s$, for the Type II, $\ell=3$, LDPC codes. 
\subsection{Relation to finite geometry codes}
The codes that result from this $\ell = 3$ construction correspond to the two-dimensional projective-geometry-based LDPC (PG LDPC) codes of \cite{xu}. With a little modification of the Type II construction, we can also obtain the two-dimensional Euclidean-geometry-based LDPC codes of \cite{xu}. \\
\indent Since a two-dimensional Euclidean geometry may be obtained by deleting certain points and line(s) of a two-dimensional projective geometry, the  graph of a two-dimensional EG-LDPC code \cite{xu} may be obtained by performing the following operations on the Type II, $\ell = 3$, graph:
\begin{enumerate}
\item In the tree $T$, the root node along with its neighbors, i.e., the constraint nodes in layer $L_1$, are deleted.
\item Consequently, the edges from the constraint nodes $B_0,\ldots,B_{p^s}$ to layer $L_2$ are also deleted.

\item At this stage, the remaining variable nodes have degree $p^s$, and 
the remaining constraint nodes have degree $p^s+1$. Now, a constraint node 
from layer $L_3$ is chosen, say, constraint node $A_{0,0}$. This node and 
its neighboring variable nodes and the edges incident on them are deleted. 
Doing so removes exactly one variable node from each class of $L_2$, and 
the degrees of the remaining constraint nodes in $L_3$ are lessened by 
one. Thus, the resulting graph is now $p^s$-regular with a girth of six, 
has $p^{2s}-1$ constraint and variable nodes , and corresponds to the 
two-dimensional Euclidean-geometry-based LDPC code $EG(2,p^s)$ of 
\cite{xu}.

\end{enumerate}

\subsection{$\ell = 4$}
\begin{enumerate}
\item The tree $T$ is now enumerated for four layers, with the nodes
  in $L_0, L_1,$ and $L_2$ labeled as in the $\ell = 3$ case. For $i =
  0,\ldots, p^s$, the constraint nodes in the $i$th class of $L_3$ are
  labeled
  $B_{i,0,0},B_{i,0,1},\ldots,B_{i,0,p^s-1},B_{i,1,0},B_{i,1,1},\ldots,B_{i,1,p^s-1}$,
  $\dots$, $B_{i,p^s-1,0},\ldots, B_{i,p^s-1,p^s-1}$.
\item The $p^{3s}$ variable nodes in the final layer $L_4$ are labeled
  $A_{0,0,0}, A_{0,0,1},\dots, A_{0,0,p^s-1}, A_{0,1,0},A_{0,1,1},
  \dots, A_{0,1,p^s-1}$, $\dots
  A_{p^s-1,0,0},A_{p^s-1,0,1},\dots,A_{p^s-1,0,p^s-1}$, $\dots,
  A_{p^s-1,p^s-1,0}, A_{p^s-1,p^s-1,1}, \dots, A_{p^s-1,p^s-1,p^s-1}$.
\item For $0\le i\le p^s-1$, $0\le j\le p^s-1$, connect the variable
  node $B_{0,i,j}$, that is in the $0^{th}$ class of $L_3$, to the
  constraint nodes $A_{i,j,0}, A_{i,j,1},\dots, A_{i,j,p^s-1}.$
\item Let $x_t(i,k) = M^{(i-1)}(k,t)$, the $(k,t)^{th}$ entry of
  $M^{(i-1)}$, and let $y_t(i,j)=M^{(i)}(j,t)$, the $(j,t)^{th}$ entry
  of $M^{(i*)}$, where $i*=i \mbox{ mod } p^s$.
\item Then, for $1\le i\le p^s$, $0\le j,k\le p^s-1$, connect the
  variable node $B_{i,j,k}$ to the constraint nodes
  \[\hspace{-0.5in}A_{0,x_0(i,k),y_0(j,k)}, A_{1,x_1(i,k),y_1(j,k)},
  \dots, A_{p^s-1,x_{p^s-1}(i,k),y_{p^s-1}(j,k)}.\]
\end{enumerate}

\begin{figure}

  \centering{\resizebox{3.25in}{1.6in}{\includegraphics{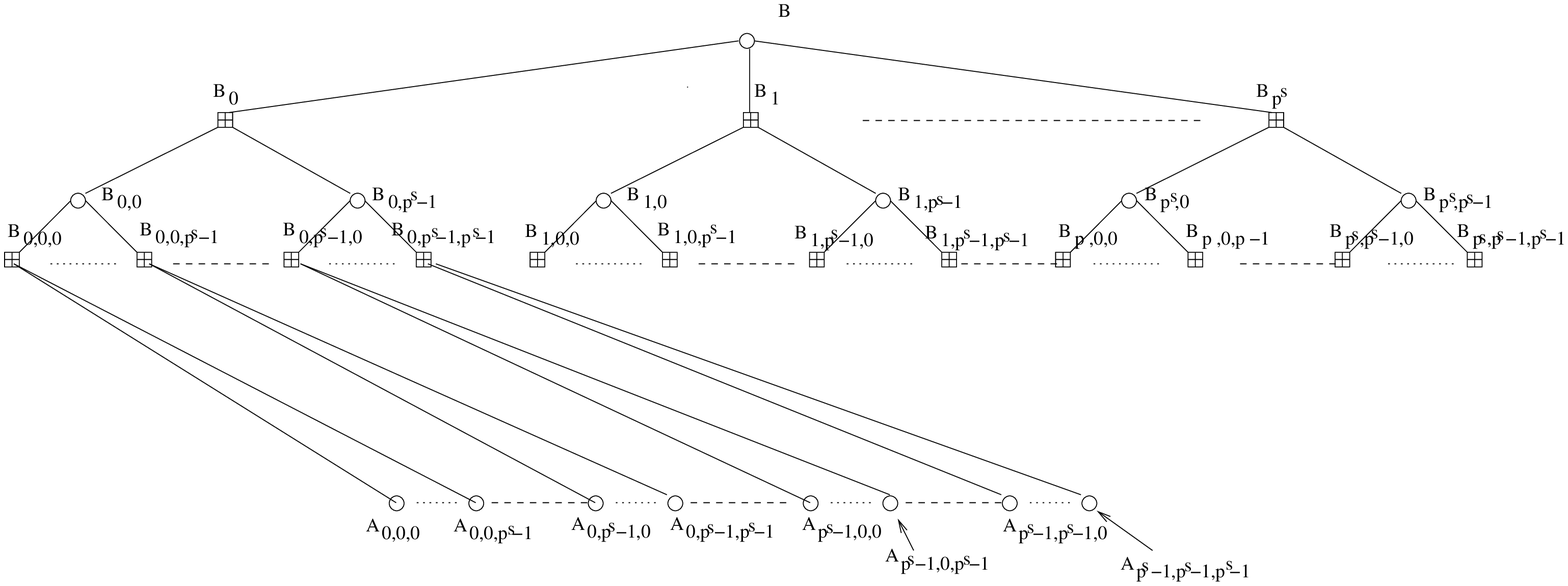}}}
\caption{Tree construction of girth 8 Type II ($\ell=4$) LDPC code.}
\label{type2g8_const}
\end{figure}

\begin{figure}

  \centering{\resizebox{3.25in}{1.6in}{\includegraphics{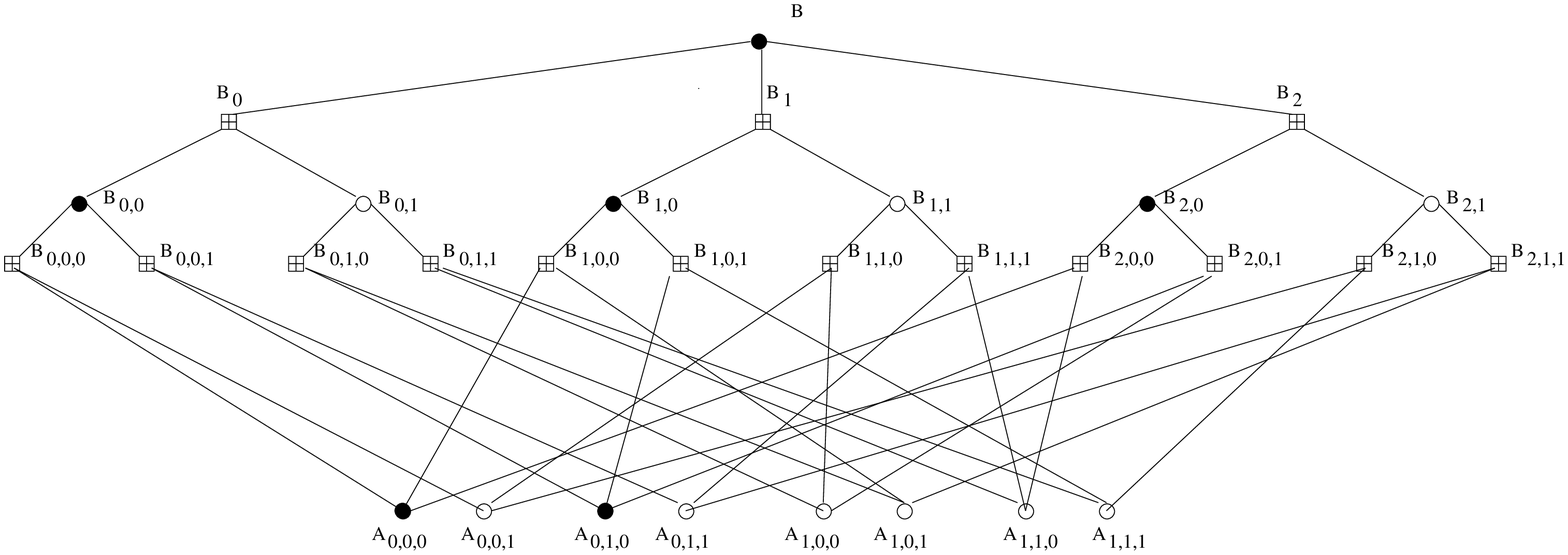}}}
\caption{Type II LDPC constraint graph having degree $d=3$ and girth $g=8$. (Shaded nodes highlight a minimum weight codeword.)}
\label{type2g8d3_graph}
\end{figure}

The  Type II, $\ell = 4$, LDPC codes have girth eight, minimum distance $d_{\min}\ge 2(p^s+1)$, and blocklength $N=1+p^s+p^{2s}+p^{3s}$. 
(We believe that the tree bound on the minimum distance is actually met
for all the Type II, $\ell=4$, codes, i.e. $d_{\min}=w_{\min}=2(p^s+1)$.)
Figure~\ref{type2g8_const} illustrates the general construction
procedure. For $d=3$, the Type II, $\ell=4$, LDPC constraint graph as
shown in Figure~\ref{type2g8d3_graph} corresponds to the
$(2,2)$-Finite-Generalized-Quadrangles-based LDPC (FGQ LDPC) code of
\cite{pascal_fgq}; the squares used for constructing this code are
\[\tiny 
\hspace{-0in} M^{(0)}=\left[\begin{array}{cc}
    0&0\\
    1&1
\end{array}\right], \ M^{(1)}=\left[\begin{array}{cc}
0&1\\
1&0
\end{array}\right].\]
We believe that the Type II, $\ell=4$, construction results in other FGQ
LDPC codes for other choices of $d$. The Type II construction
algorithm can be modified for larger $\ell$ by involving more
iterations of the MOLS in the connection scheme, as will be discussed
in a forthcoming paper.

\section{Simulation Results}
\begin{figure}
            \centering
          {
          \resizebox{3in}{2.5in}{\includegraphics{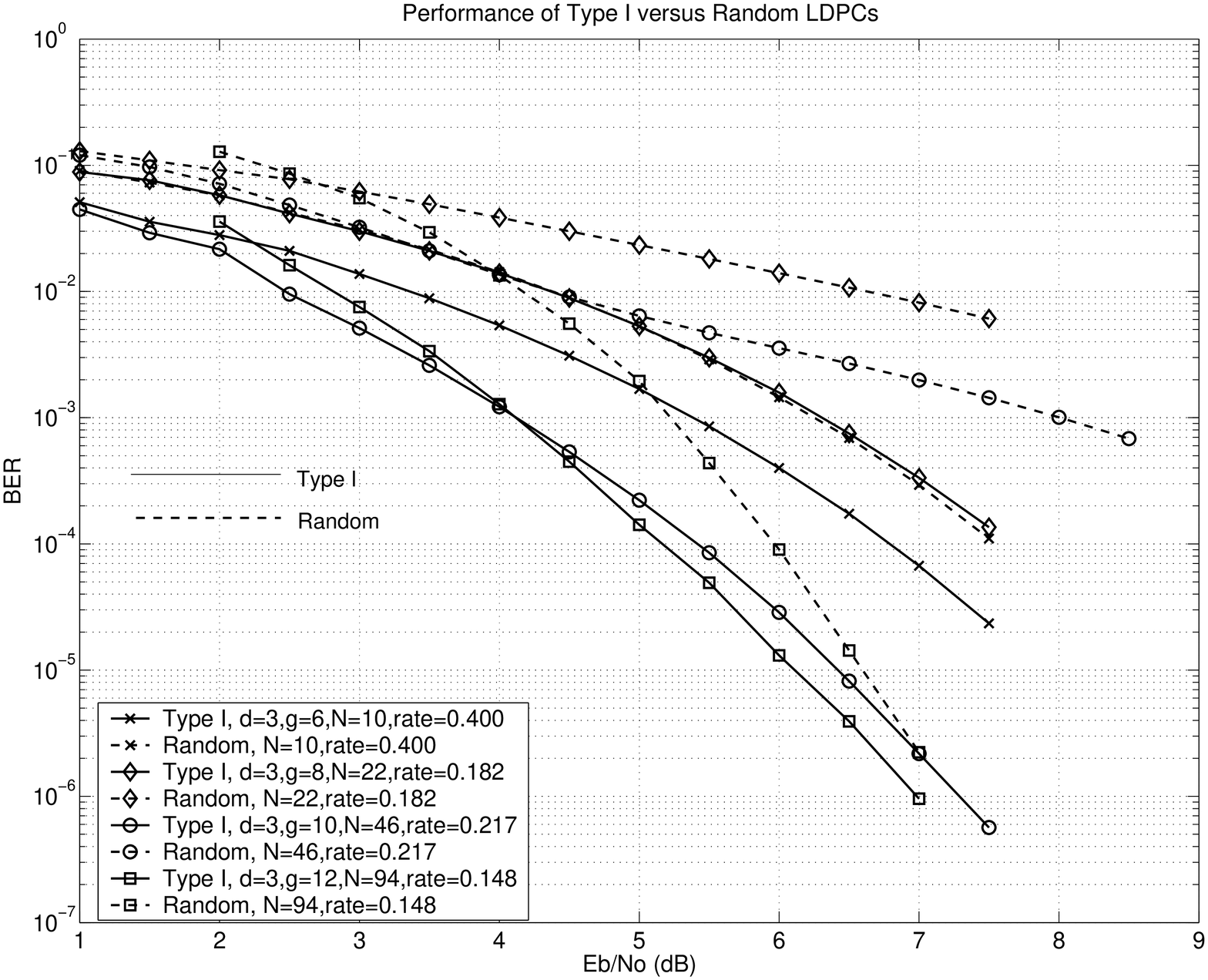}}}
          \caption{Performance of Type I-A versus Random LDPC codes with min-sum iterative decoding.}
\label{type1_perf}
\end{figure}

\begin{figure}
            \centering{
            \resizebox{3in}{2.5in}{\includegraphics{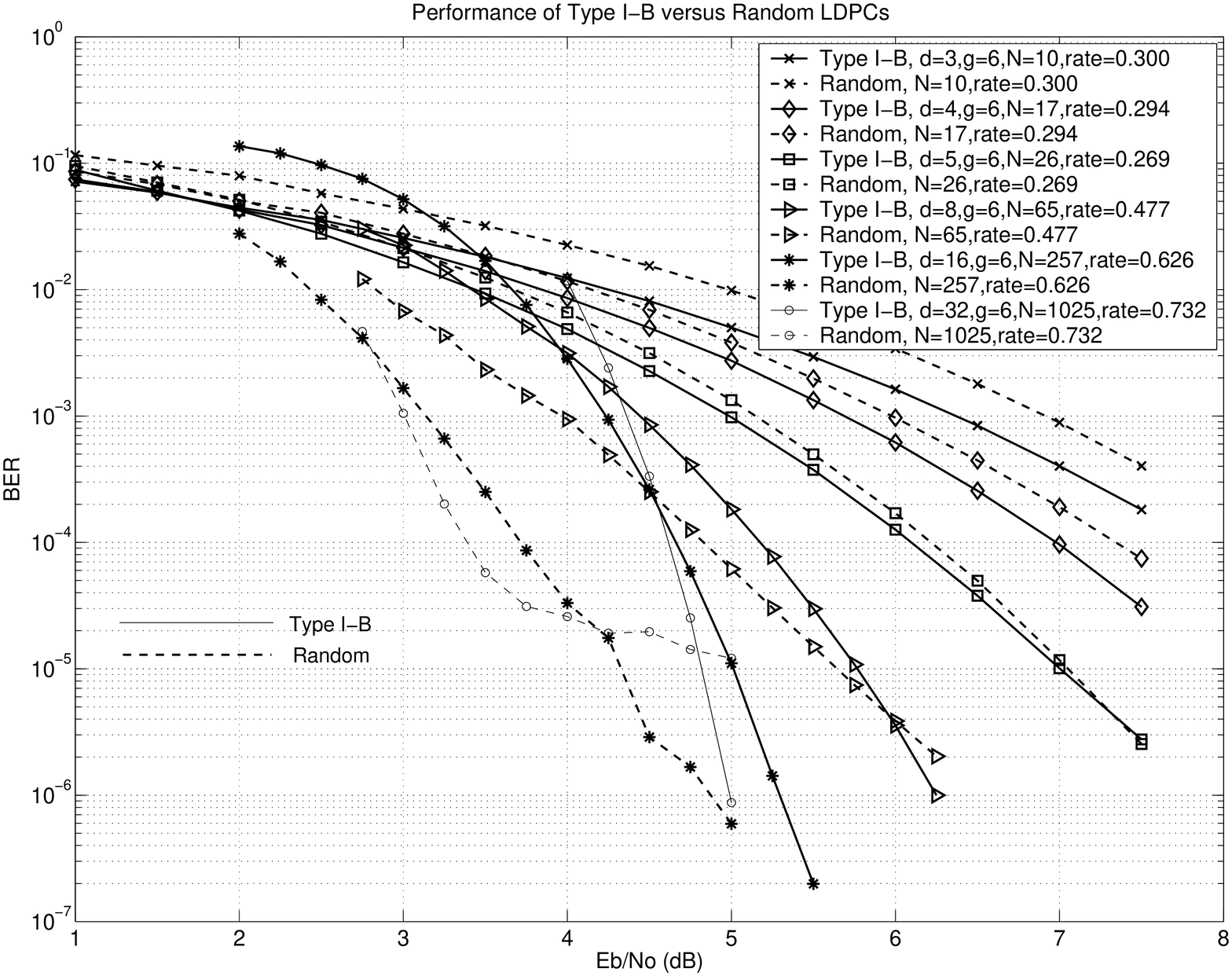}}}
            \caption{Performance of Type I-B versus Random LDPC codes with min-sum iterative decoding.}
\label{type1B_perf}
\end{figure}

\begin{figure}
            \centering
          {
          \resizebox{3in}{2.5in}{\includegraphics{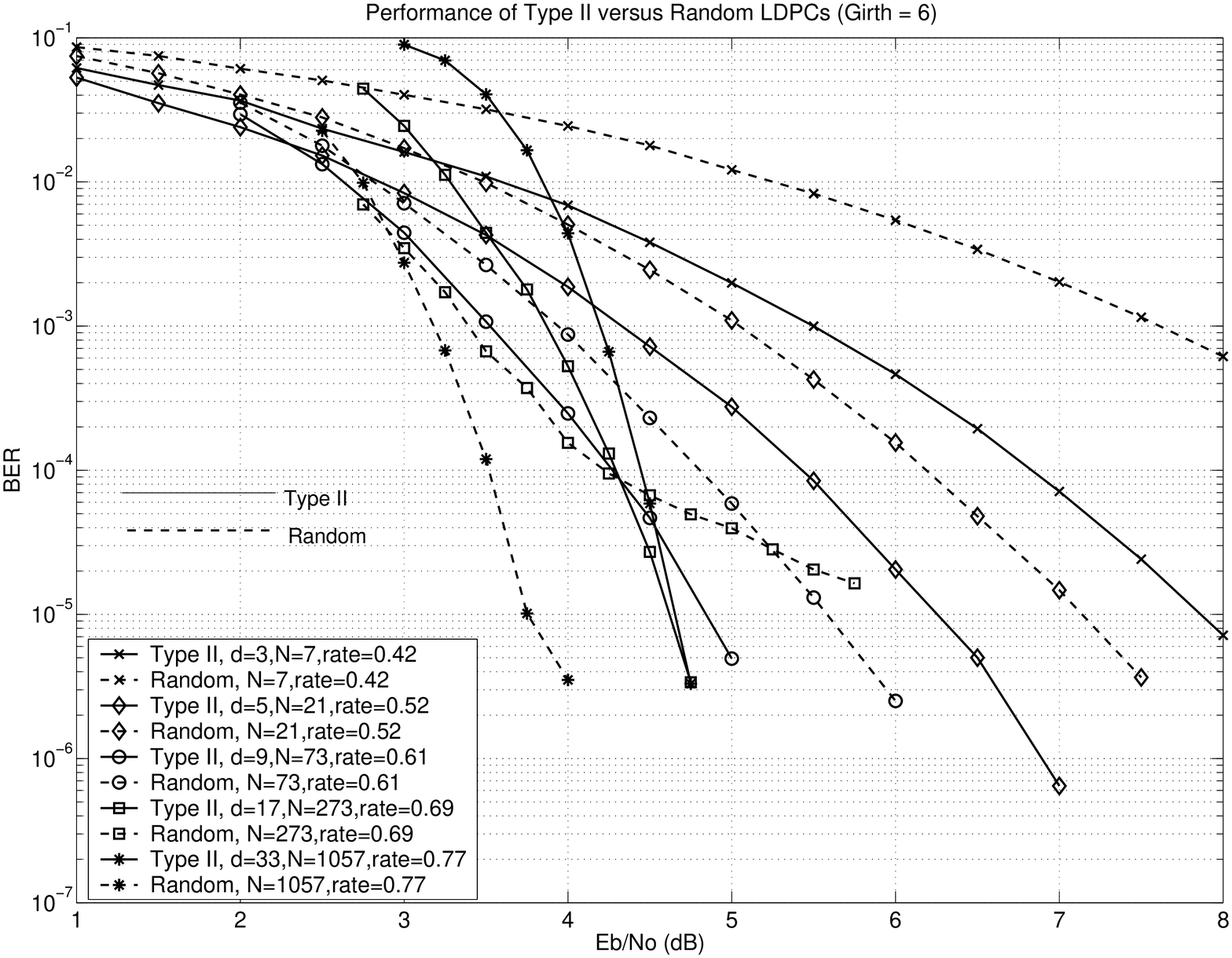}}}
          \caption{Performance of girth 6 Type II versus Random LDPC codes with min-sum iterative decoding.}
\label{type2g6_perf}
\end{figure}

\begin{figure}
            \centering{
            \resizebox{3in}{2.5in}{\includegraphics{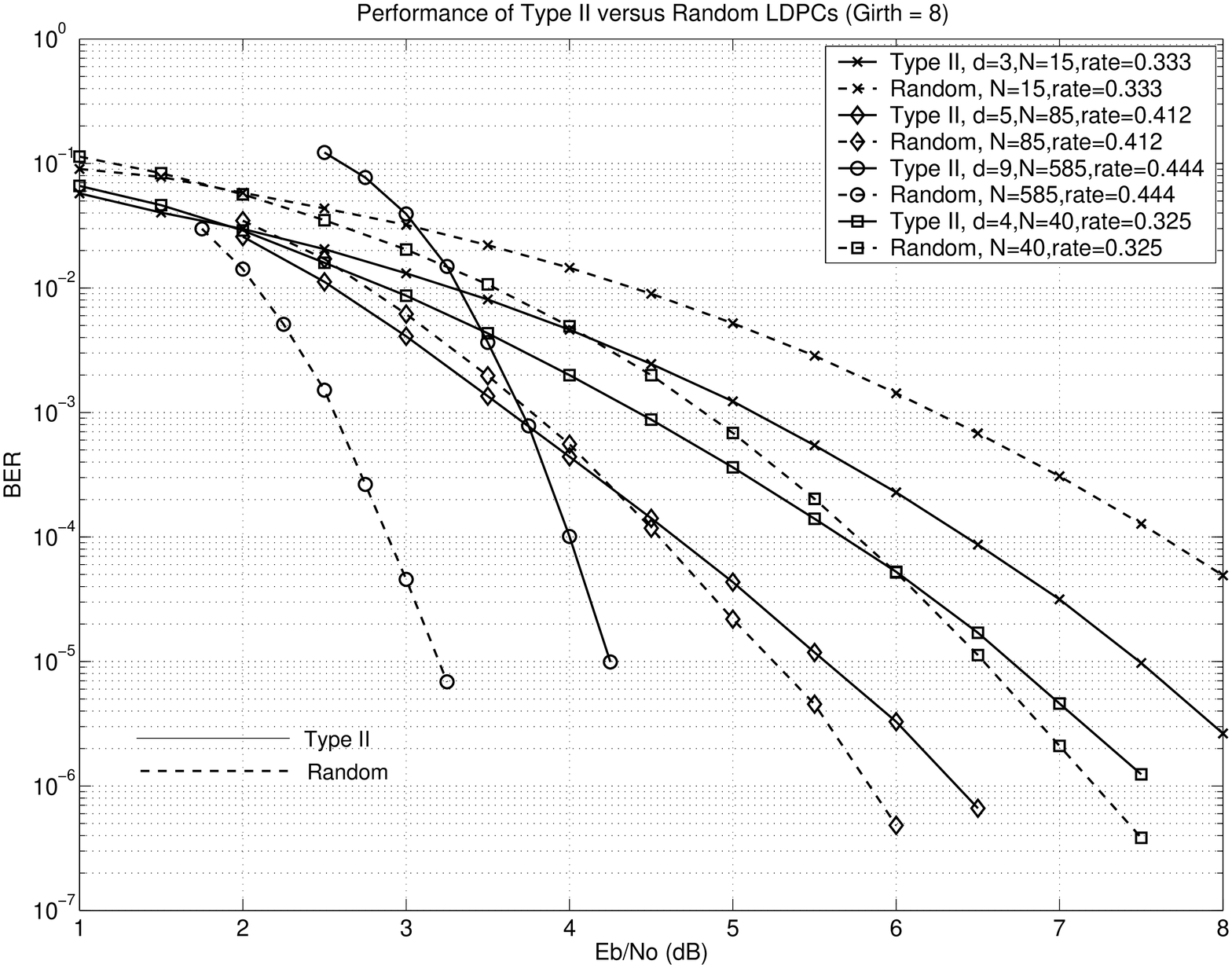}}}
            \caption{Performance of girth 8 Type II versus Random LDPC codes with min-sum iterative decoding.}
\label{type2g8_perf}
\end{figure}

Figures \ref{type1_perf}, \ref{type1B_perf}, \ref{type2g6_perf}, \ref{type2g8_perf} show the bit-error-rate performance of Type I-A, Type I-B, Type II girth six, and Type II girth eight LDPC codes, respectively over a binary input additive white Gaussian noise channel with min-sum iterative decoding. The performance of regular or semi-regular randomly constructed LDPC codes of comparable rates and blocklengths are also shown. (All of the random LDPC codes compared in this paper have a variable node degree of three and are constructed from the online LDPC software available at 

{\tt \scriptsize http://www.cs.toronto.edu/$\tilde{}$ radford/ldpc.software.html}.) 

Figure \ref{type1_perf} shows that the Type I-A LDPC codes perform substantially better than their random counterparts.  Figure \ref{type1B_perf} reveals that the Type I-B LDPC codes perform better than comparable random LDPC codes at short blocklengths; but as the blocklengths increase, the random LDPC codes tend to perform better in the waterfall region. Eventually however, as the SNR increases, the Type I-B LDPC codes outperform the random ones since, unlike the random codes, they do not have a prominent error floor. Figure \ref{type2g6_perf} reveals that the performance of Type II girth-six LDPC codes is also significantly better than comparable random codes; these codes correspond to the two dimensional PG LDPC codes of \cite{xu}. Figure \ref{type2g8_perf} indicates the performance of Type II girth-eight LDPC codes; these codes perform comparably to random codes at short blocklengths, but at large blocklengths, the random codes perform better as they have larger relative minimum distances compared to the Type II girth-eight LDPC codes.

As a general observation, min-sum iterative decoding of most of the tree-based LDPC codes (particularly, Type I-A, Type II,  and some Type I-B) presented here did not typically reveal detected errors, i.e., errors caused due to the decoder failing to converge to any valid codeword within the maximum specified number of iterations. Detected errors are caused primarily due to the presence of pseudocodewords, especially those of minimal weight. We think that the lack of detected errors with iterative decoding of these LDPC codes is primarily due to their good minimum pseudocodeword weight $w_{\min}$. 

\section{Conclusions}
The Type I construction yields a family of LDPC codes that, to the best of our knowledge, do not correspond to any of the LDPC codes obtained from finite geometries or other geometrical objects. The two tree-based constructions presented in this paper yield a wide range of codes that perform well when decoded iteratively, largely due to the maximized minimal pseudocodeword weight. However, the overall minimum distance of the code is relatively small. Constructing codes with larger minimum distance, while still maintaining $d_{\min} = w_{\min}$, remains an open problem.






\end{document}